# Hyper Vision Net: Kidney Tumor Segmentation Using Coordinate Convolutional Layer and Attention Unit

D.Sabarinathan[1], M.Parisa Beham[1], and S.M.Md.Mansoor Roomi[2]

[1]Couger Inc, Tokyo, Japan
[2]Sethu Institute of Technology, Virudhunagar, Tamilnadu, India
[3]Thiagarajar college of Engineering, Madurai, Tamil Nadu, India

**Abstract.** KiTs19 challenge paves the way to haste the improvement of solid kidney tumor semantic segmentation methodologies. Accurate segmentation of kidney tumor in computer tomography (CT) images is a challenging task due to the non-uniform motion, similar appearance and various shape. Inspired by this fact, in this manuscript, we present a novel kidney tumor segmentation method using deep learning network termed as Hyper vision Net model. All the existing U-net models are using a modified version of U-net to segment the kidney tumor region. In the proposed architecture, we introduced supervision layers in the decoder part, and it refines even minimal regions in the output. A dataset consists of real arterial phase abdominal CT scans of 300 patients, including 45964 images has been provided from KiTs19 for training and validation of the proposed model. Compared with the state-of-the-art segmentation methods, the results demonstrate the superiority of our approach on training dice value score of 0.9552 and 0.9633 in tumor region and kidney region, respectively.

## 1 Introduction

As per the study of American Cancer Society, Kidney cancer is among the 10 most common cancers in both men and women. Overall, the lifetime risk for evolving kidney cancer in men is about 1 in 48 and for women is 1 in 83. Many kidney cancers are found at an early stage but others are found at a more advanced stage. One of the main reasons for this is, the kidneys are deep inside the body so that small tumors cannot be seen during a physical exam [1]. Wide range of imaging techniques have been enabled to detect the kidney tumors at the early stage. In recent years, this approach has become more popular because it offers a chance to remove only the tumor lesion while preserving the much healthy kidney parenchyma. This method can also be useful for small renal masses treatment. After accurate segmentation of kidney tumor, some valuable information like renal volume, anatomy of kidney, etc., can be obtained. According to the clinical study [2], it is very difficult to predict the locations of various renal tumors in CT or MRI medical images. Moreover, these renal tumors are having dissimilar shapes and sizes, and most of the tumors have a similar appearance with their parenchyma and other nearby tissues. Thus the segmentation of kidney tumor region become a very challenging task in the CT images.

In the literature several algorithms and networks have been introduced to segment renal tumor from CT images. M G Lingararu et al [3] proposed a computer-aided clinical tool based on adaptive level sets which was used to analyze 125 renal lesions from contrast-enhanced abdominal CT scans of 43 patients. In this method tumors are robustly segmented with 0.80 overlap between manual and semi-automated quantifications. The method also identified morphological discrepancies among various kinds of lesions. Lee et al [4] presented an automated method for detecting and segmenting small renal mass (SRM) in contrast-enhanced CT images using texture and context feature classification. Their experimental results produced specificity of 99.63% in the SRM detection.

B Shah et al [5] presented a computer aided segmentation technique using machine learning algorithms. Guanyu Yang et al [6] presented an automated kidney segmentation in CT images based on multi atlas image registration. First they registered the down-sampled patient image with a set of low-resolution atlas images and thus the left and right kidneys are segmented. In the next step, the kidneys are cropped and aligned with the set of high-resolution atlas images to obtain final segmented results.

Recently more papers have been published on renal tumor segmentation using deep convolutional networks. In [7], a patch-wise approach is used to train the ConvNet to predict the class membership of the central voxel in 2D patches. Then densely processing the ConvNet over each slice of a CT scan produced a segmented kidney tumors. Skalski et al [8] presented a novel hybrid level set method with elliptical shape constraints for kidney segmentation. Here RUSBoost and the decision trees technique were used to identify the kidney and tumor. This approach resolved the main problems like class imbalance and the number of voxels required to classify. Their proposed model produced an overall accuracy amounts to 92.1%. In [9] automated segmentation of kidneys using fully convolutional neural networks is presented, which is trained end-to-end, on slice wise axial-CT sections. Similarly Wang et al [10] proposed an interactive segmentation by incorporating CNNs into a bounding box. They have also made the CNN model more adaptive by image specific fine tuning process.

Even though CNNs have achieved benchmark performance for automatic medical image segmentation, accuracy and robustness is still a challenging issue. To address these problems, U-Net is proposed [11] for automatic medical image segmentation where the U network synthesize the significant information by minimizing a cost function in the first half of the network and construct an image at the second half. Inspired by the U-Net model, we approached the present task of kidney tumor segmentation by proposing a novel Hyper vision Net model, in which we introduced Hyper vision layers in the decoder part of the U-net architecture and thus it refines even very small regions in the output tumor image. Our proposed framework segment the tumor regions of the kidney accurately and the results are proved quantitatively as well as qualitatively.

## 2   Dataset

A collection of multi-phase CT images, segmenting masks and their respective clinical reports of 300 patients have been provided in the challenging KiTs19 dataset [12]. For training, 210 (70%) of these patients CT images were selected randomly and released publicly for the 2019 MICCAI Kidney Tumor Segmentation (KiTS) Challenge. The objective of this dataset is to create a reliable learning-based kidney and kidney tumor semantic segmentation methods. Among the 210 patient's CT images, 32175 and 13790 images are used for the purpose of training and validation respectively.  Fig.1 shows an example of CT images of 5 patients.

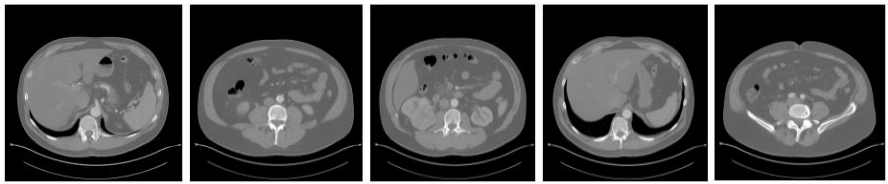

**Fig.1.** An example of CT scan images from KiTs19 Challenge dataset

## 3   Proposed Method

In this section, the detailed structure of the proposed Hyper vision net model and the modified loss function is described.

### 3.1   Image Preprocessing

In this work, before training the model, all the CT images are resized into 256 x 256 and divided by the value 255 to normalize the pixel value between 0 to 1.

### 3.2   Hyper Vision Net Model

Fig. 2 shows the detailed architecture of the proposed Hyper vision Net model. The network has the properties of encoder and decoder structure of vanilla U-Net [13]. As proposed by [14], firstly the input image is passed into the coordinate convolution layer and then it is passed into the encoder part of the Hyper vision Net model. Here, to improvise the generalization capacity of the model, the coordinate convolutional layer is used which helps the network to select the features related to translation invariance. Also, with the support of this layer, spatial coordinates are mapped with the Cartesian coordinate's space through the use of extra coordinate channels. This kind of mapping provides power to the model to use either complete or varying degree of translation features. As suggested in the original paper [14], we have also used two extra coordinate channels (i, j) in the proposed work.

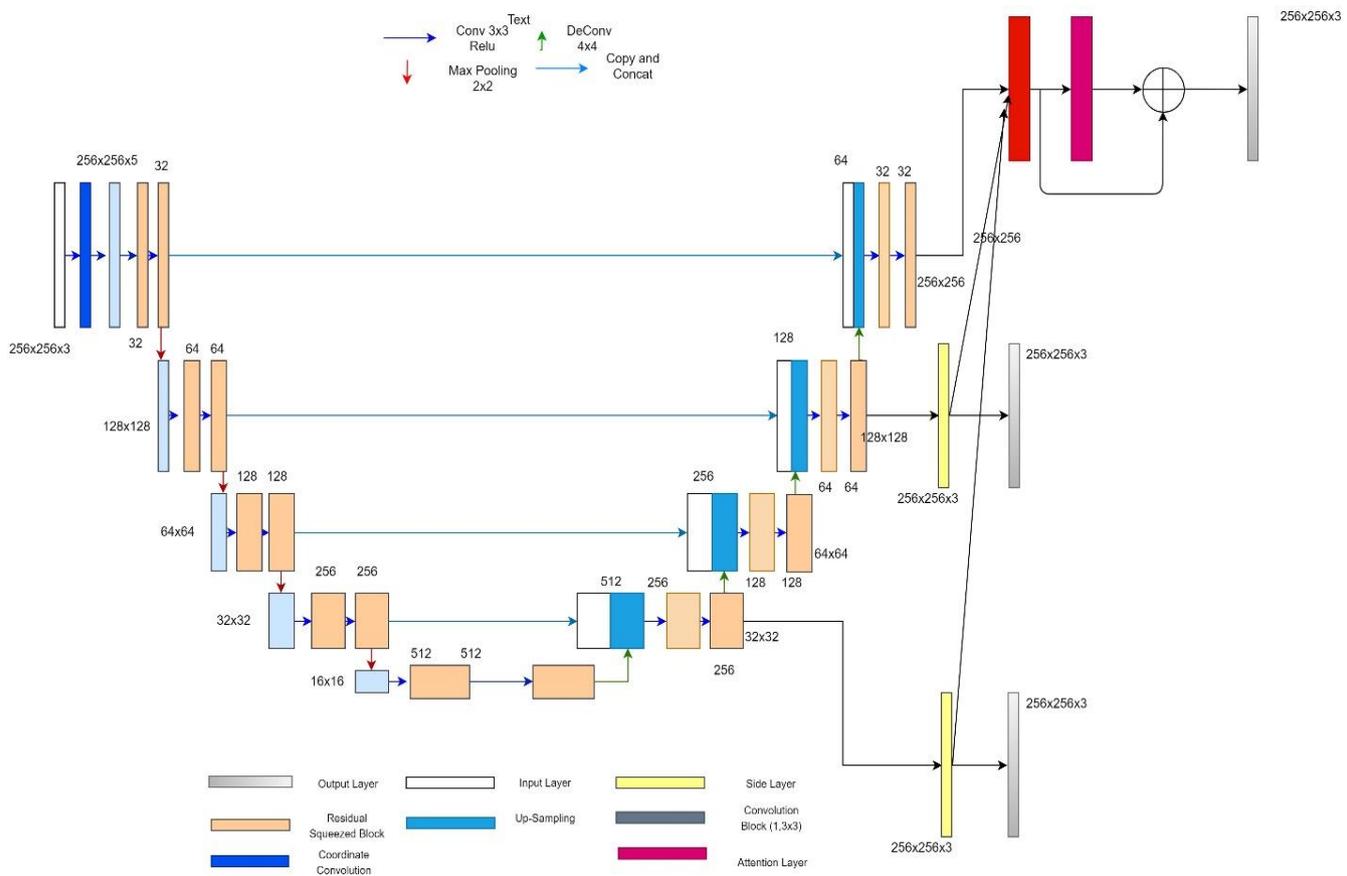

**Fig.2.** An overview of detailed architecture of Hyper vision Net

During down-sampling in the encoder phase, there are four blocks have been used. In each block, the first layer is a 3×3 convolutional layer, followed by two residual blocks, and at the end, a 2×2 max-pooling layer is added. Our network also contains a deep residual block where the details of the block is shown in Fig. 3. We increased the depth of the residual block[17] to acquire multiple features, and in turn it enhances the performance of tumor segmentation. Similar to encoder phase, the decoder phase is also utilized the same blocks except the max-pooling layers replaced with the up-sampling layers. To get more robust feature maps after down-sampling, we added one 3x3 convolutional layer and residual block before passing the features into the encoder phase. Batch normalization and dropout layers are incorporated in both encoder and decoder phase. The network is trained with the Exponential Linear activation function.

Inspired by Skeleton Network [15], two Hyper Vision Layers are introduced in the decoder part which is used to refine even very small regions in the output. The output of two decoder layers is then fused with final encoder output. This fused output was given to the attention layer [16]. For the given intermediate feature map, the proposed model successively infers attention maps along channel and spatial, a two separate dimensions, then the attention maps are multiplied with the input feature map for adaptive feature refinement. The output of the attention layer is added with fused output. Further, the output of Hyper vision layer and the final layer are passed through a soft max layer individually under the supervision of ground truth. The final output layer has three channels such as, tumor region, kidney region and the background.

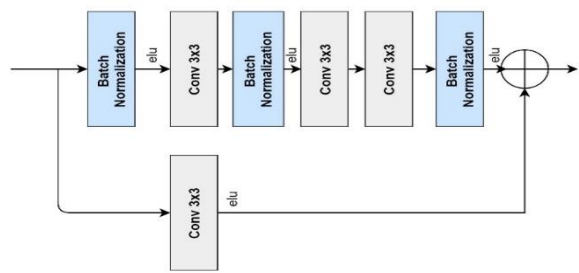

**Fig.3**. Residual Block used in the Hyper Vision Net

### 3.3 Loss Function

In this work, Adam[18] optimizer is applied which perfectly update network weights in an iterative manner in training data. Adam makes average in first moment but also in second moments of gradients to adapt the learning rate parameter. Our Loss function is the sum of categorical cross entropy Dice loss channel one($C_0$) and Dice Loss channel two ($C_1$) as defined in eqn. (1).

$$Loss = L + DiceLoss(C_0) + DiceLoss(C_1) \quad (1)$$

$$DiceLoss = 1 - \frac{2\sum_{n=k}^{i=0} y_i p_i + \epsilon}{\sum_{n=k}^{i=0} y_i + \sum_{n=k}^{i=0} p_i + \epsilon} \quad (2)$$

$$L = -\sum_{j=0}^{M} \sum_{i=0}^{N} y_{ij} \log p_{ij} \quad (3)$$

$L$ is the cross entropy loss. In eqn. (2) $y_i$ and $p_i$ are the ground truth and the predicted segmented images respectively. Also, to ensure the loss function stability the coefficient $\epsilon$ is used.

## 4   Experiment and Results

### 4.1   Training

The proposed network is trained with three outputs, which include the two hyper vision layers and one fused output layer. The weight updates performed with Adam optimizer using a learning rate of 0.001 and reduced after ten epochs to 10% if there is no improvement in the validation loss. The batch size is chosen to four, and the total epochs are set to 500. The model is trained using Nvidia 1080 GTX GPU.

### 4.2   Result and Discussion

To evaluate the performance of the proposed Hyper vision Net model, the standard Dice score is considered as an evaluation metric. For our experimentation, we were provided with 32175 and 13790 images as a training and validation images respectively. Table 1 shows the segmentation results of the proposed Hyper vision Net model for training and validation images with and without considering the attention layer. From the table it is observed that, during training, the proposed method achieves the Dice score of 0.8967 and 0.9535 for the tumor region and kidney region by involving attention layer in the proposed network respectively.  Similarly, our network with attention layer achieves Dice score of 0.8967 and 0.9535 for tumor and kidney region respectively. It is also inferred that, without incorporating attention layer in the Hyper vision Net, Dice score of  training and validation experimentation is reduced to 0.8186 and 0.84 respectively for tumor region, whereas for kidney region it is only 0.9375 and 0.93. From the experimental results we understand the power of attention layer in the proposed network. Attention layer helps the network to focus more in the segmentation region and utilize the fused features well.

**Table 1.** Results of the proposed Hyper vision Net model on KiTs19 dataset with and without attention layer

| Model Type | Image Type | Number of Images | Dice score of Tumor Region | Dice score of Kidney Region |
|---|---|---|---|---|
| Without Attention Layer | Training | 32175 | 0.8186 | 0.9375 |
| | Validation | 13790 | 0.84 | 0.93 |
| With Attention Layer | Training | 32175 | **0.9552** | **0.9633** |
| | Validation | 13790 | **0.8967** | **0.9535** |

The qualitative results of KiTs19 dataset on our proposed Hyper vision net is shown in Fig.3. Fig.3 (a) and Fig.3 (b) shows the input images and their respective ground truth images used for experimentation. The segmented output images are depicted in Fig.3(c). In the output image the blue colored spot is a tumor region, whereas the red color spot is the kidney region. Total background other than the tumor and kidney regions are neglected for easy interpretation. From the qualitative results it is observed that the final segmented output is merely similar to the ground truth images which shows the efficacy of our proposed Hyper vision Network.

## 5    Conclusion

Wide range of accurate image segmentation techniques are required to detect the kidney tumors at the early stage. Motivated by the superior performance of Convolutional Neural Networks, in this paper, a Hyper vision Net architecture is presented to segment the kidney and tumor region which is automatic and accurate. This challenge is carried out using KiTs19 dataset. The performance of our method is reported quantitatively and qualitatively for the given type of training and validation images. Our method achieved a maximum Dice score of 0.9633 for the training set and 0.9535 for the validation set. Comparatively the proposed Hyper vision Net reported best segmentation results in terms of Dice score.

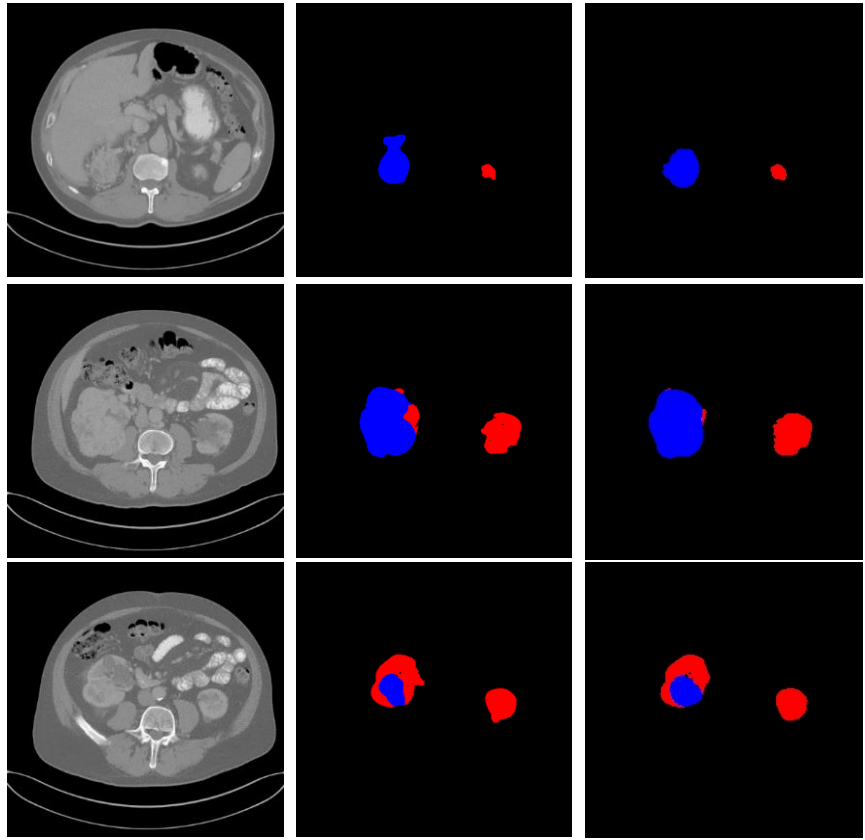

Fig.3. Illustration of original input CT images and their respective Kidney and tumor segmented output images